\documentstyle[11pt,paspconf,epsfig]{article}
\def\HII{\hbox{H~$\scriptstyle\rm II\ $}}

\def\kmsmpc{\,{\rm km\,s^{-1}\,Mpc^{-1}}}

\def\ldunits{\,{\rm h_{50}\,ergs\,s^{-1}\,Hz^{-1}\,Mpc^{-3}}}
\def\Haunits{\,{\rm ergs\,s^{-1}}}
\def\lunits{\,{\rm ergs\,s^{-1}\,Hz^{-1}}}
\def\msun{\,{\rm M_\odot}}
\def\mden{\,{\rm h_{50}^2\,M_\odot\,Mpc^{-3}}}
\def\mdden{\,{\rm M_\odot\,Mpc^{-3}}}
\def\sfrd{\,{\rm M_\odot\,yr^{-1}\,Mpc^{-3}}}

\def\etal{{et al.\ }}
\def\ub{U-B}

\def\vi{V-I}

\def\spose#1{\hbox to 0pt{#1\hss}}
\def\lta{\mathrel{\spose{\lower 3pt\hbox{$\mathchar"218$}}
     \raise 2.0pt\hbox{$\mathchar"13C$}}}
\def\gta{\mathrel{\spose{\lower 3pt\hbox{$\mathchar"218$}}
     \raise 2.0pt\hbox{$\mathchar"13E$}}}
 
\begin{document}

\title{The Evolution of Luminous Matter in the Universe}
\author{Piero Madau}
\affil{Space Telescope Science Institute, 3700 San Martin Drive, Baltimore,
MD 21218}

\begin{abstract}
I review a technique for interpreting faint galaxy data which traces the 
evolution with cosmic time of the galaxy luminosity density, as 
determined from several deep spectroscopic samples and the {\it Hubble Deep 
Field} imaging survey. The method relies on the rest-frame UV and 
near-IR continua of galaxies as indicators, for a given initial mass function
(IMF) and dust content, of their instantaneous star formation rate (SFR) and 
total stellar mass, and offers the prospect of addressing in a coherent 
framework an important set of subjects: cosmic star formation history, dust 
in primeval galaxies, shape of the IMF, stellar mass-to-light ratios 
of present-day galaxies, extragalactic background light, Type~II supernovae
and heavy element enrichment history of the universe.  The global 
spectrophotometric 
properties of field galaxies are well fit by a simple stellar evolution model,
defined by a time-dependent SFR per unit comoving volume, a universal IMF
which is relatively rich in massive stars, and a modest amount of dust 
reddening. The model is able to account for the entire background 
light recorded in the galaxy counts down to the very faint magnitude levels 
probed by the HDF, and produces visible mass-to-light ratios at 
the present epoch which are consistent with the values observed in nearby 
galaxies of various morphological types. The bulk ($\gta 60\%$) of the stars 
present today formed relatively recently ($z\lta 1.5$), consistently with 
the expectations from a broad class of hierarchical clustering cosmologies, and
in good agreement with the low level of metal enrichment observed at high
redshifts in the damped Lyman-$\alpha$ systems. Throughout this review I 
emphasize how the poorly constrained amount of starlight that was absorbed by 
dust and reradiated in the far-IR at early epochs represents one of the biggest
uncertainties in our understanding of the evolution of luminous matter in the 
universe. A ``monolithic collapse'' model, where half of the present-day 
stars formed at $z>2.5$ and 
were enshrouded by dust, can be made consistent with the global history of 
light, but overpredicts the metal mass density at high redshifts as 
sampled by QSO absorbers. 

\end{abstract}

\section{Introduction}

As the best view to date of the optical sky at faint flux levels, the {\it 
Hubble Deep Field} (HDF) has offered to many astronomers the opportunity to 
study the galaxy population in unprecedented details. In particular, the deep 
HDF images have rapidly become a key testing ground for the two competing 
scenarios that 
have been widely used in the past few years to interpret the observed 
properties of galaxies. In what may be called the ``traditional'' scheme, one 
starts from 
the local measurements of the distribution of galaxies as a function of 
luminosity and Hubble type and models their photometric evolution assuming a 
well defined collapse epoch, pure-luminosity evolution thereafter, and a set 
of parameterized star formation histories (Tinsley 1980; Bruzual \& Kron 1980;
Koo 1985; Pozzetti, Bruzual, \& Zamorani 1996). These, together with an 
initial mass function (IMF) and a world model, are then adjusted to match the 
observed number counts, colors, and redshift distributions. Beyond its
intrinsic simplicity, the main advantage of this kind of approach is that it 
can be made consistent with the classical view that ellipticals 
and spiral galaxy bulges formed early in a single burst of duration 1 Gyr or 
less (e.g., Bower \etal 1992; Ortolani \etal 1995). Because in this 
``monolithic'' collapse scenario for spheroids much of the action happens at 
high-$z$, however, these models predict, in the absence of a significant amount
of dust obscuration, far more Lyman-break ``blue dropouts''
than are seen in the HDF (Pozzetti \etal 1997; Ferguson \& Babul 1997). 
Moreover, they cannot reproduce the rapid evolution 
-- largely driven by late-type galaxies -- of the optical luminosity density 
with lookback time observed by Lilly \etal (1996) and Ellis \etal (1996). 

A more physically motivated way to interpret the observations is to construct
semianalytic hierarchical models of galaxy formation and evolution (White \&
Frenk 1991; Kauffmann, White, \& Guiderdoni 1993; Cole \etal 1994; Baugh \etal 
1997). Here, one starts from a power spectrum of primordial density 
fluctuations, 
follows the growth of dark matter halos by accretion and mergers, and 
adopts various prescriptions for gas cooling, star formation, feedback from
supernovae and stellar winds, and dynamical friction. These are tuned to 
match the statistical properties of 
both nearby and distant galaxies.  In this scenario, there is no period when 
bulges and ellipticals form rapidly as single units and are very bright: 
rather, small objects form first, eventually settling into disks, and merge 
continually over a range of redshifts to make ellipticals (see reviews by G. 
Kauffmann and S. White in these proceedings). While reasonably successful 
in recovering the counts, colors, and redshift 
distributions of faint galaxies, a generic difficulty of such models is the 
inability to simultaneously reproduce the observed local luminosity density 
and the zero-point of the Tully-Fisher relation (White \& Frenk 1991; Cole 
\etal 1994).

In this talk I will describe an alternative method, which focuses on the 
emission properties of the galaxy population {\it as a whole}. It traces the 
cosmic evolution with redshift of the galaxy luminosity density -- as 
determined from several deep spectroscopic samples and the HDF imaging 
survey -- and offers the prospect of an empirical determination of the global 
star formation history of the universe and
initial mass function of stars independently, e.g., of the merging
histories and complex evolutionary phases of individual galaxies. The technique
relies on two basic properties of stellar populations: a) the
UV-continuum emission in all but the oldest galaxies is dominated by
short-lived massive stars, and is therefore a direct measure, for a given IMF
and dust content, of the instantaneous star formation rate (SFR); and b) the 
rest-frame near-IR
light is dominated by near-solar mass evolved stars that make up the bulk of a
galaxy's visible mass, and can then be used as a tracer of the integrated 
stellar mass density. By modeling the ``emission history'' of the universe at 
ultraviolet, optical, and near-infrared wavelengths from the present epoch to 
$z\approx 4$, I will try to shed some light on what I believe are key questions
in structure formation and evolution studies, and provide a first glimpse to 
the history of the conversion of neutral gas into stars within galaxies.

The initial applications of this novel technique have been presented by 
Lilly \etal (1996), Madau \etal (1996, hereafter M96), and Madau, Pozzetti, 
\& Dickinson (1997). A complementary effort -- which starts 
instead from the analysis of the evolving gas content and metallicity of the 
universe -- can be found in Fall, Charlot, and Pei (1996). A flat cosmology 
with $q_0=0.5$ and $H_0=50\,h_{50}\kmsmpc$ will be adopted in this review.

\begin{figure}[b!]
\centerline{\epsfig{file=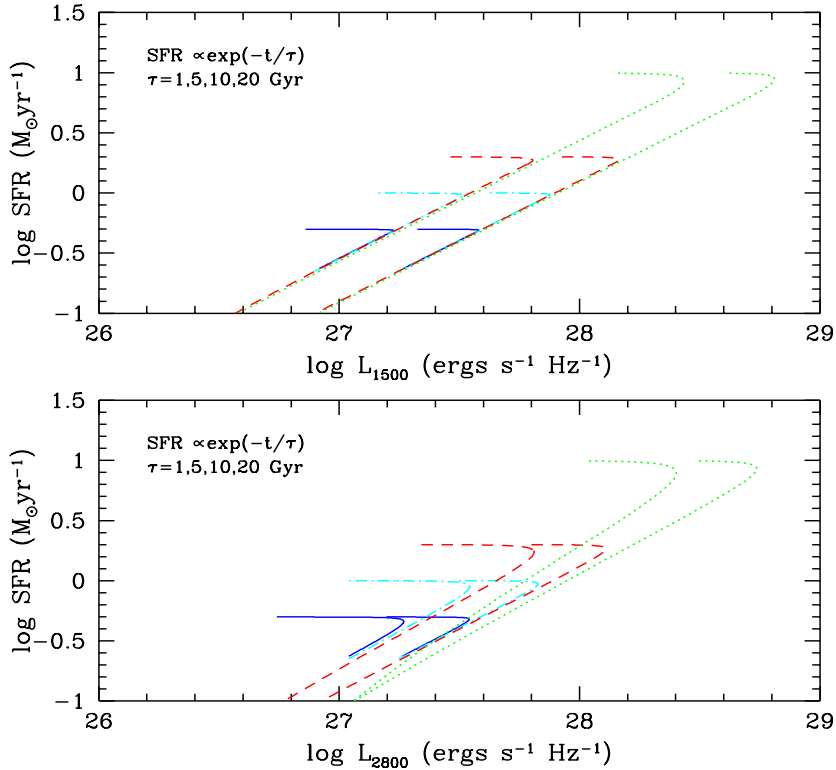,height=4.3in,width=4.5in}}
\caption{SFR-UV relation for models with various exponentially
declining star formation rates at ages between 0.01 and 15 Gyr. {\it Solid
lines:} $\tau=20$ Gyr. {\it Dot-dashed lines:} $\tau=10$ Gyr. {\it Dashed
lines:} $\tau=5$ Gyr. {\it Dotted lines:} $\tau=1$ Gyr. The set of curves on
the left-hand side of the plot assume a Scalo IMF, the ones on the right-hand
side a Salpeter function.
\label{fig1}}
\end{figure}

\section{Indicators of Past and Present Star Formation Activity}

Stellar population synthesis has become a standard technique to study the
spectrophotometric properties of galaxies. In the following, I shall make 
extensive use of the latest version of Bruzual \& Charlot (1993) isochrone 
synthesis code, optimized with an updated library of stellar spectra 
(Bruzual \& Charlot 1997), to predict the time change of the spectral energy 
distribution of a stellar population. I will consider here two possibilities 
for the IMF, a Salpeter (1955) function and a Scalo (1986) IMF,  
which is flatter for low-mass stars and significantly less rich in massive 
stars than Salpeter. In all models the metallicity is fixed to solar values 
and the IMF is truncated at 0.1 and 125 $\msun$. 

\subsection{Birthrate--Ultraviolet Relation}

The UV continuum emission from a galaxy with significant ongoing star formation
is entirely dominated by late-O/early-B stars on the main sequence. As these
have masses $\gta 10\msun$ and lifetimes $t_{MS}\lta 2\times 10^7\,$yr, the
measured luminosity becomes proportional to the stellar birthrate and
independent of the galaxy history for $t\gg t_{MS}$. This is depicted in Figure
1, where the power radiated at 1500 \AA\ and 2800 \AA\ is plotted against the
instantaneous SFR for a model stellar population with different star formation
laws, SFR $\propto \exp(-t/\tau)$, where $\tau$ is the duration of the burst.
After an initial transient phase where the UV flux rises rapidly and the
turnoff mass drops below 10$\msun$, a steady state is reached where one can
write 
\begin{equation}
L_{UV}={\rm const}\times {{\rm SFR}\over \msun\, {\rm yr}^{-1}}\, \lunits,
\end{equation}
with const$=(3.5\times 10^{27}, 5.0\times 10^{27})$ at (1500\AA, 2800\AA) for a
Scalo IMF, and const$=8.0\times 10^{27}$ in the same wavelength range 
for a Salpeter IMF, quite insensitive to the details of the past star formation
history. 
Note how, for burst durations $\lta 1 $Gyr and a Scalo IMF, the luminosity at
2800 \AA\ becomes a poor SFR indicator after a few $e$-folding times, when the
contribution of intermediate-mass stars becomes significant. After averaging
over the whole galaxy population, however, we will find that the (unreddened)
UV continuum is always a good tracer of the instantaneous rate of conversion 
of cold gas into stars. 

\subsection{Birthrate--H{\bf$\alpha$} Relation}

In the optical wavelength range, the H$\alpha$($\lambda6563$) luminosity from 
a galaxy is another probe of its current star formation activity (Kennicutt 
1983), as, for ionization bounded \HII regions, the integrated line emission 
scales directly with the Lyman-continuum flux of the embedded OB stars. 
Assuming case-B recombination theory and a Salpeter function, the H$\alpha$ 
luminosity can be related to the local SFR according to 
\begin{equation}
L_{\rm H\alpha}=1.5\times 10^{41}\left({{\rm SFR} \over\msun\,{\rm yr}^{-1}}
\right)\, \Haunits.
\end{equation} 
The coefficient for a Scalo IMF is approximately four times smaller.
Note that, shortward of the Lyman edge, the differences in the predicted 
ionizing radiation from model atmospheres of hot stars can be quite large 
(Charlot 1996a), and must be taken into account when interpreting the results 
of surveys for H$\alpha$-emitting galaxies (Gallego \etal 1995). 

\subsection{Supernova Frequency}

The frequency of supernovae (SNe) is also intimately related, for a given IMF,
to the stellar birthrate. This is true, in particular, for ``core-collapse 
supernovae'', SN~II and SN~Ib/c, which have massive, short-lived ($t_{MS}\lta 
2\times 10^7$ yr) progenitors. It might also be approximately 
true for Type Ia SNe -- believed to result from the thermonuclear 
disruption of accreting CO white dwarfs in binary systems (for a recent review
see Ruiz-Lapuente, Canal, \& Burkert 1997) --  if the favourite 
route follows a fast evolutionary track, and the deflagration occurs within few
Gyr of (binary) stellar birth. Since the Hubble time is equal to 3 Gyr 
at $z=1.7$, 
it would then be possible to neglect the delay between stellar birth and the 
SN~Ia it eventually yields at all epochs $z\ll 1.7$ (but see Yungelson \& 
Livio 1997).

For a Salpeter IMF and a lower mass cutoff for the progenitor star of 10$\msun$,
the core-collapse supernova rate (SNR) can be related to the stellar birthrate 
according to   
\begin{equation}
{\rm SNR}=0.0055\times \left({{\rm SFR} \over\msun\, {\rm yr}^{-1}} \right)
\, {\rm yr}^{-1}.
\end{equation} 
The coefficient for a Scalo IMF is 2.6 times smaller. I will show later
how SNe may provide an independent test of the global star 
formation history of the universe at low redshifts.

\subsection{Mass--Infrared Relation}

If we assume assume a time-independent IMF, we can use the results of 
stellar
population synthesis modeling, together with the observed UV emissivity, to 
infer the evolution of the star formation activity in the universe (M96). The 
biggest uncertainty in this procedure is due to dust reddening, as newly formed
stars which are completely hidden by dust would not contribute to the UV 
luminosity. The effect is potentially more serious at high redshifts, as for
a fixed observer-frame bandpass, one is looking further in the ultraviolet 
with increasing lookback time. For example, SMC-type foreground dust with 
$E(B-V)\gta 0.1$ would produce an extinction at 1500~\AA\  greater than 1.3 
magnitudes. On the other hand, it should be possible to test the hypothesis 
that star formation regions remain largely unobscured by dust throughout much 
of galaxy evolution by looking at the 
near-infrared light density. This will be affected by dust only in the most 
extreme, rare cases, as it takes an $E(B-V)>4$ mag to produce an optical depth 
of unity at 2.2 \micron. 

\begin{figure}[b!]
\centerline{\epsfig{file=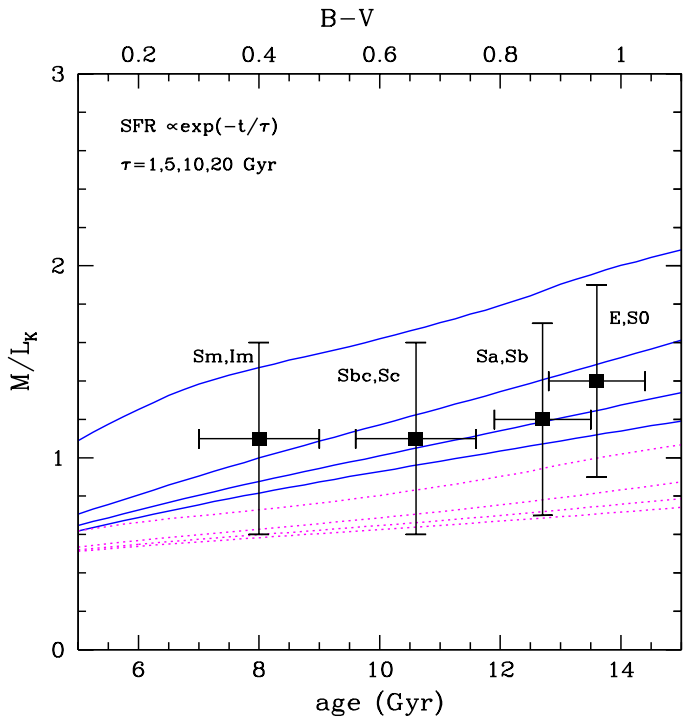,height=4.3in,width=4.5in}}
\caption{Total (processed gas+stars) mass-to-$K$ band light ratio versus age
for models with various exponentially declining star formation rates. {\it
Solid lines:} Salpeter IMF. {\it Dotted lines:} Scalo IMF. From top to bottom,
each set of curves depict the values for $\tau=1, 5, 10,$ and 20 Gyr,
respectively. The data points show the luminous mass-to-infrared light ratio
versus $B-V$ color (top axis) observed in nearby galaxies of various
morphological types (see Charlot 1996b). The observations refer to the mass
within the galaxy Holmberg radius, where the contribution by the dark matter
halo is expected to be small.
\label{fig2}}
\end{figure}

Although different types of stars -- such as supergiants, AGB, and red
giants -- dominate the $K$-band emission at different ages in an evolving
stellar population, the mass-to-infrared light ratio is relatively insensitive
to the star formation history (Charlot 1996b). Figure 2 shows $M/L_K$ (in solar
units) as a function of age for models with various exponentially declining
SFR compared to the values observed in nearby galaxies of early to late 
morphological types. As the stellar population ages, the mass-to-infrared light
ratio remains very close to unity, independent of the galaxy color and 
Hubble type. We can use this interesting property to estimate the visible mass
in galaxies from the local $K-$band luminosity density, $\log 
\rho_K(0)=27.05\pm0.1\ldunits$ (Gardner \etal 1997). The observed range
$0.6\,h_{50}\lta M/L_K\lta 1.9\,h_{50}$ translates into a mass density of 
stars$+$gas at the present day of 
\begin{equation} 
2\times 10^8 \lta \rho_{s+g}(0) \lta 6\times 10^8 \mden
\end{equation}
($0.003\lta \Omega_{s+g}\lta 0.009$). I will show later how the observed 
integrated UV emission, with the addition of some modest amount of reddening, 
may account for the bulk of the baryons traced by the $K$-band light, and how 
initial mass functions with relatively few high-mass stars (such as the Scalo 
IMF), or models with a large amount of dust extinction at all epochs will 
tend to overproduce the near-infrared emissivity. 

\section{Galaxy Emissivity as a Function of Redshift}

The integrated light radiated per unit volume from the entire galaxy population
is an average over cosmic time of the stochastic, possibly short-lived star
formation episodes of individual galaxies, and should follow a relatively simple
dependence on redshift. In the UV -- where it is proportional to the global SFR
-- its evolution should provide information on the mechanisms which
may prevent the gas within virialized dark matter halos from radiatively cooling
and turning into stars at early times, or on the epoch when galaxies exhausted 
their reservoirs of cold gas. From a comparison between different wavebands it 
should be possible to set constraints on the average IMF and dust content of
galaxies. 

The observed continuum emissivity, $\rho_\nu(z)$, from the present 
epoch to $z\approx 4$ is shown in Figure 3 in six broad passbands centered 
around 0.15, 0.2, 0.28, 0.44, 1.0, and 2.2 \micron.  The data are taken 
from the $K$-selected redshift survey of Gardner \etal (1997) and Cowie \etal
(1996), the $I$-selected  CFRS 
(Lilly \etal 1996) and $B$-selected Autofib (Ellis \etal 1996) samples,
a redshift survey of galaxies imaged in the rest-frame ultraviolet at
2000 \AA\ with the FOCA balloon-borne camera (Treyer \etal 1997), the 
photometric redshift catalog for the HDF of Connolly \etal (1997) -- which 
take advantage of deep infrared observations by Dickinson \etal (1997) --,
and the color-selected UV and blue ``dropouts'' of M96 (see also Madau 1997).
They have all (except for the Treyer \etal value) been corrected for 
incompleteness by integrating over the best-fit Schechter function in each 
redshift bin, 
\begin{equation}
\rho_\nu(z)=\int L_\nu\phi(L_\nu,z)dL_\nu=\Gamma(2+\alpha)\phi_*L_*. 
\end{equation}
As it is not possible to reliably determine the faint end slope of the 
luminosity function from the Connolly \etal (1997) and M96 data sets, a value of
$\alpha=-1.3$ has been assumed at each redshift interval for comparison with
the CFRS sample (Lilly \etal 1995). The error bars are typically less than 0.2
in the log, and reflect the uncertainties present in these corrections and, in
the HDF $z>2$ sample, in the volume normalization and color-selection region.

\begin{figure}[b!]
\centerline{\epsfig{file=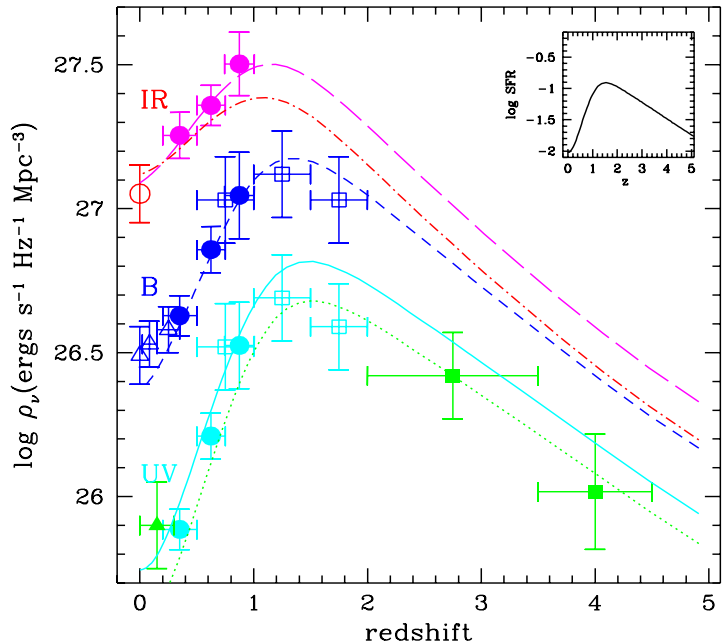,height=4.3in,width=4.5in}}
\caption{Evolution of the comoving luminosity density at rest-frame 
wavelengths of 0.15 ({\it dotted line}), 0.28 ({\it solid line}), 0.44 
({\it short-dashed line}), 1.0 ({\it long-dashed line}), and 2.2 ({\it 
dot-dashed line}) \micron. The data points with error bars are taken from 
Lilly \etal (1996) ({\it filled dots}), Connolly \etal (1997) ({\it empty 
squares}), Madau \etal (1996) and Madau
(1997) ({\it filled squares}), Ellis \etal (1996) ({\it empty triangles}),
Treyer \etal (1997) ({\it filled triangle}), and Gardner \etal (1997) ({\it 
empty dot}). The inset in the upper-right corner of the plot shows the SFR 
density ($\sfrd$) versus redshift which was used as input to the population 
synthesis code. The model assumes a Salpeter IMF, SMC-type dust in a 
foreground screen, and a universal $E(B-V)=0.1$.
\label{fig3}}
\end{figure}

Despite the obvious caveats due to the likely incompleteness in the data sets,
different selection criteria, and existence of systematic uncertainties in the
photometric redshift technique, the spectroscopic, photometric, and Lyman-break
galaxy samples appear to provide a remarkably consistent picture of the
emission history of field galaxies.\footnote{While there is no evidence for a
gross mismatch at the $z\approx 2$ transition between the photometric redshift
sample of Connolly \etal (1997) and the M96 UV dropout sample, one should note
that only one spectroscopic confirmation has been obtained so far at  $z\approx
4$ (Dickinson 1997).}~ This points to a rapid drop in the
volume-averaged SFR in the last 8--10 Gyr, and to a redshift range where the 
bulk of the stellar population was actually assembled: $1\lta z\lta 2$. 

\section{Population Synthesis}

It is interesting to see now whether a simple stellar evolution model, defined
by a time-dependent SFR per unit volume and a universal IMF, may reproduce the
global UV, optical, and near-IR photometric properties of galaxies. In a 
stellar system with arbitrary star formation rate, the luminosity density at 
time $t$ is given by the convolution integral 
\begin{equation}
\rho_\nu(t)=\int^t_0 L_\nu(\tau)\times {\rm SFR}(t-\tau)d\tau, 
\label{eq:rho} 
\end{equation}
where $L_\nu(\tau)$ is the specific luminosity radiated per unit initial mass
by a generation of stars with age $\tau$. After relating the observed UV 
continuum emissivity to a SFR density, one can then use, e.g.,
Bruzual-Charlot's synthesis code to predict the time evolution of the
spectrophotometric properties of a stellar population in a comoving volume large
enough to be representative of the universe as a whole. In doing so, we will
bypass all the ambiguities associated with the study of morphologically
distinct samples, but, at the same time, we will not be able to specifically
address the evolution of particular subclasses of objects, like the oldest
ellipticals, whose star formation histories may have differed significantly
from the global average. It is fair at this stage to point out two other
significant limitations of this approach: a) It focuses on the emission 
properties of ``normal'', optically-selected 
field galaxies which are only moderately affected by dust -- a typical spiral 
emits 30\% of its energy in the far-infrared region (Saunders \etal 1990).
Starlight which is completely blocked from view even in the near-IR
by a large optical depth in dust will not be recorded by this technique, and the
associated baryonic mass and metals missed from our census. The contribution of
infrared-selected dusty starbursts to the present-day total stellar mass density
cannot be very large, however, for otherwise the current limits to the energy
density of the mid- and far-infrared background would be violated (Puget \etal
1996; Kashlinsky, Mather, \& Odenwald 1996; Fall \etal 1996;
Guiderdoni \etal 1997). Locally, infrared luminous galaxies are known to
produce only a small fraction of the IR luminosity of the universe (Soifer \&
Neugebauer 1991); and b) It does not include the effects of cosmic chemical 
evolution on the predicted galaxy colors. All the population synthesis models 
assume solar metallicity, and thus will generate colors that are slightly too
red for objects with low metallicity, e.g. truly primeval galaxies.

\subsection{A Fiducial Model: Salpeter IMF}

Figure 3 shows the model predictions for the evolution of $\rho_\nu$ at
rest-frame ultraviolet to near-infrared frequencies for a Salpeter IMF.  
In the absence of dust reddening, this relatively flat IMF generates spectra 
that are slightly too blue to reproduce the observed mean (luminosity-weighted over
the entire population) galaxy colors. The effect of dust attenuation can 
be included by multiplying equation (\ref{eq:rho})
by $p_{\rm esc}$, a redshift-independent term equal to the fraction of
emitted photons which are not absorbed by dust. For purposes of illustration,
I will assume a foreground screen model, $p_{\rm esc}=\exp(-\tau_\nu)$, and
SMC-type dust.\footnote{Since what is relevant here is the absorption opacity,
I have multiplied the extinction optical depth by a factor of 0.6, as the
albedo of dust grains is known to approach asymptotically 0.4--0.5 at
ultraviolet wavelengths (see, e.g., Pei 1992).}~ This should only be 
regarded as a crude approximation, since hot stars can be heavily embedded in dust within
star-forming regions, there will be variety of extinction laws, and the dust 
content of galaxies will evolve with redshift. While the existing data are too
sparse to warrant a more elaborate analysis, this simple prescription will 
highlight the main features and assumptions of the model. 
The shape of the predicted and observed $\rho_\nu(z)$ relations is then found
to agree better 
to within the uncertainties if a modest amount of dust extinction, $E(B-V)=0.1$,
is included.  In this case, the observed UV luminosities must be corrected 
upwards by a factor of 1.4 at 2800 \AA\, and 2.1 at 1500 \AA. 

As expected, while the
ultraviolet emissivity traces remarkably well the rise, peak, and sharp drop in
the instantaneous star formation rate (the smooth function shown in the inset
on the upper-right corner of the figure), an increasingly large component of
the longer wavelengths light reflects the past star formation history. The peak
in the luminosity density at 1.0 and 2.2 \micron\ occurs then at later epochs,
while the decline from $z\approx 1$ to $z=0$ is more gentle than observed at
shorter wavelengths. In the instantaneous recycling approximation (Tinsley 
1980), the total stellar mass density produced at time $t$ is 
\begin{equation}
\rho_s(t)=(1-R)\int_0^t {\rm SFR}(t)dt, 
\end{equation}
where $R$ is the mass fraction of a generation of stars that is returned to the
interstellar medium, $R\approx 0.3$ for a Salpeter IMF ($R$ is closer to 0.2 fora Scalo function). The total stellar mass density at $z=0$ is then
$\rho_s(0)=3.7\times 10^8\mdden$, with a fraction close to 65\% being produced
at $z>1$, and only 20\% at $z>2$.  In the assumed
cosmology, about half of the stars observed today are more than 9 Gyr old, and
only 20\% are younger than 5 Gyr.

\begin{figure}
\centerline{\epsfig{file=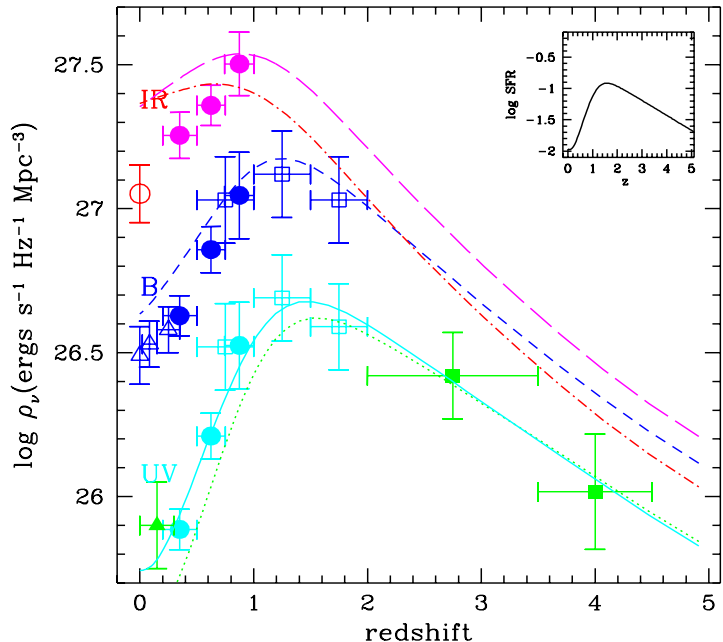,height=4.3in,width=4.5in}}
\caption{Same as in Figure 3, but assuming a Scalo IMF and no dust extinction.
This model overproduces the local $K$-band emissivity by a factor of 2.
\label{fig4}}
\end{figure}

\subsection{A Case with a Scalo IMF}

Figure 4 shows the model predictions for a Scalo IMF. The fit to
the data is now much poorer, since this IMF generates spectra that are too red
to reproduce the observed mean galaxy colors, as first noted by Lilly \etal
(1996). Because of the relatively large number of solar mass stars formed, it
produces too much long-wavelength light by the present epoch. The addition of
dust reddening would obviously make the fit even worse. The total stellar mass
density produced is similar to the Salpeter IMF case. 

\section{Clues to Galaxy Formation and Evolution}

The results shown in the previous section have significant implications for our
understanding of the global history of star and structure formation. Here I 
discuss a few key issues which will assist in interpreting the evolution of
luminous matter in the universe. 

\subsection{The Brightness of the Night Sky}

\begin{figure}
\centerline{\epsfig{file=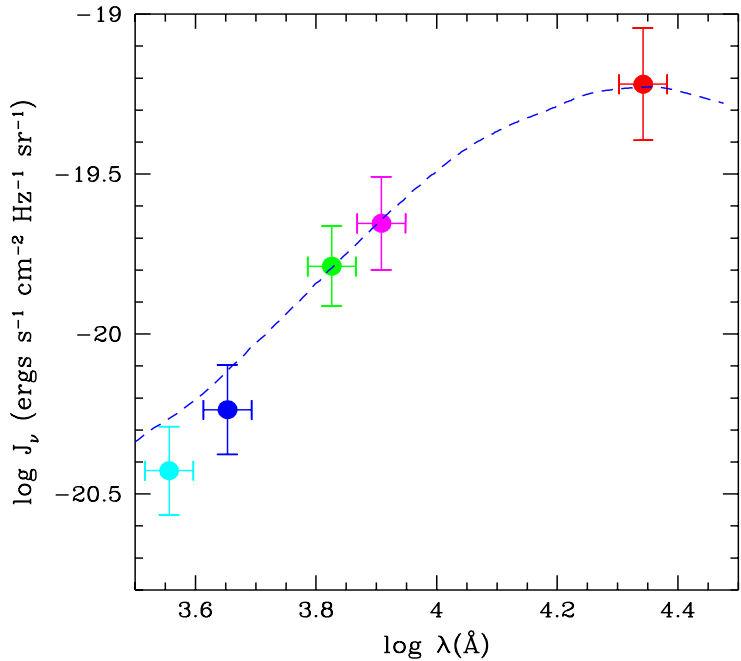,height=4.3in,width=4.5in}}
\caption{Spectrum of the extragalactic background light as derived from a 
compilation of ground-based and HDF galaxy counts (see Pozzetti \etal 1997).
The 2$\sigma$ error bars arise mostly from field-to-field variations. {\it
Dashed line:} Model predictions for the fiducial model (star formation history 
of Figure 3). 
\label{fig5}} 
\end{figure}

An important check on the inferred emission history of field galaxies
comes from a study of the extragalactic background light (EBL), an indicator of
the total optical luminosity of the universe. The contribution of known galaxies
to the EBL can be calculated directly by integrating the emitted flux times the
differential galaxy number counts down to the detection threshold. I have used
a compilation of ground-based and HDF data down to very faint magnitudes
(Pozzetti \etal 1997; Williams \etal 1996) to compute the mean surface
brightness of the night sky between 0.35 and 2.2 \micron. 
The results are plotted in Figure 5, along with the EBL spectrum
predicted by our modeling of the galaxy luminosity density, 
\begin{equation}
J_\nu={1\over 4 \pi}\int_0^\infty dz {dl\over dz}\rho_{\nu'}(z)
\end{equation}
where $\nu'=\nu(1+z)$ and $dl/dz$ is the cosmological line element. The overall
agreement is remarkably good, with the model spectrum being only slightly bluer,
by about 20--30\%, than the observed EBL. The straightforward conclusion of
this exercise is that {\it the star formation history depicted in Figure 3
appears able to account for the entire background light recorded in the
galaxy counts down to the very faint magnitudes probed by the HDF.} 

\subsection{Luminous Mass-To-Light Ratios of Present-Day Galaxies}

The fiducial Salpeter IMF model generates a present-day 
stellar mass density of $\Omega_sh_{50}^2\approx 0.004$, about 10\% of the 
nucleosynthesis constrained baryon density, $\Omega_bh_{50}^2\approx 
0.05\pm 0.01$ (Walker \etal 1991). The (luminosity-weighted) visible 
mass-to-light ratios range from about 4 in the $B$-band to 1 in $K$, consistent 
with the values observed in nearby galaxies of various morphological types
(see, e.g., Persic \& Salucci 1992 and references therein).
Note, however, that the predicted $M/L$ are quite sensitive to the
lower-mass cutoff of the IMF, as very-low mass stars can contribute
significantly to the mass but not to the integrated light of the whole stellar
population. For example, a lower cutoff of 0.2$\msun$ instead of the 
0.1$\msun$ adopted would decrease the mass-to-light ratio by a factor of 1.3.
Although one could in principle reduce the inferred star formation density by 
adopting a top-heavy IMF, richer in massive UV-producing stars, in practice a 
significant amount of dust reddening -- hence of ``hidden'' star formation -- 
would then be required to match the observed galaxy colors. The net effect 
of this operation would be a baryonic mass comparable to the estimate above
and a large infrared background. 

\begin{figure}
\centerline{\epsfig{file=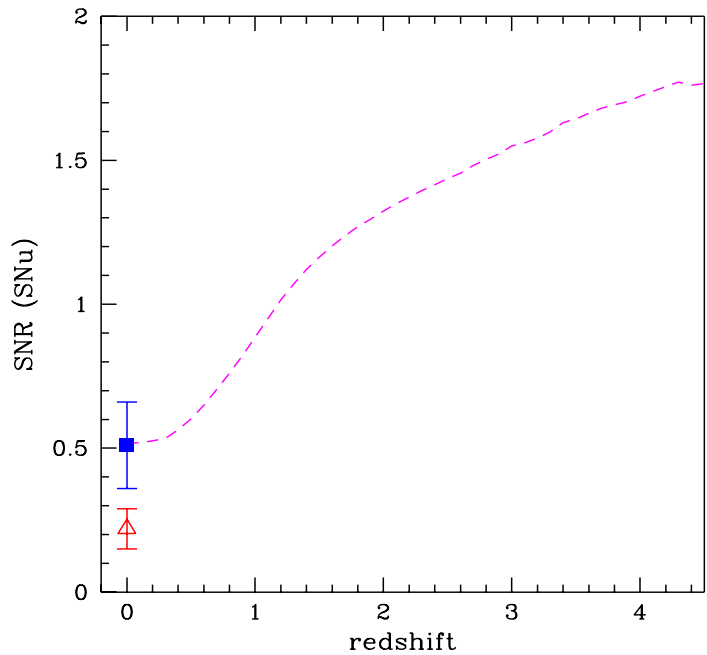,height=4.3in,width=4.5in}}
\caption{Rest-frame rate of core-collapse supernovae (Type~II$+$Ib/c) in 
SNu (SNe 
per $10^{10}L_\odot (B)$ per century) versus redshift. The data points with 
error bars are taken from van den Bergh \& Tammann (1991) ({\it filled 
square}), and Cappellaro \etal (1997) ({\it empty triangle}). They have been normalized 
according to the local blue luminosity function by spectral type of Heyl 
\etal (1997). The {\it dashed line} depicts the rate predicted from our 
Salpeter IMF, $E(B-V)=0.1$ fiducial model, assuming a lower mass cutoff for the
progenitors of 10 $M_\odot$. 
\label{fig6}}
\end{figure}

\subsection{Evolution of the Supernova Rate with Redshift}

In recent years there has been a renewed interest in the search for 
supernovae. While SN~Ia may provide one of the best distance indicators
at high redshifts (Kim \etal 1997), a direct measurement of the 
rate of Type II SNe could be used as an independent test for the 
star formation and heavy element enrichment history of the universe.  
Figure 6 shows the normalized rate of core-collapse SNe predicted by our 
best-fit model as a function of cosmic time. The frequency 
is computed in the rest-frame of the supernovae, and is expressed in 
SNu (SNe per $10^{10}L_\odot (B)$ per century). Unlike the rate per unit 
comoving volume, which will trace the rise, peak, and drop of the star 
formation density, the blue luminosity-weighted frequency is a monotonic
increasing function of redshift. The predicted rate is in good agreement 
(to within 0.2 in the log) with the local values estimated by van den Bergh 
\& Tammann (1991), and Cappellaro \etal (1997). In the near future, ongoing
searches for distant SNe should provide enough data to constrain the star
formation history of the universe at intermediate redshifts, $z\approx 0.5-1$
(Della Valle \& Madau 1997). 

\begin{figure}
\centerline{\epsfig{file=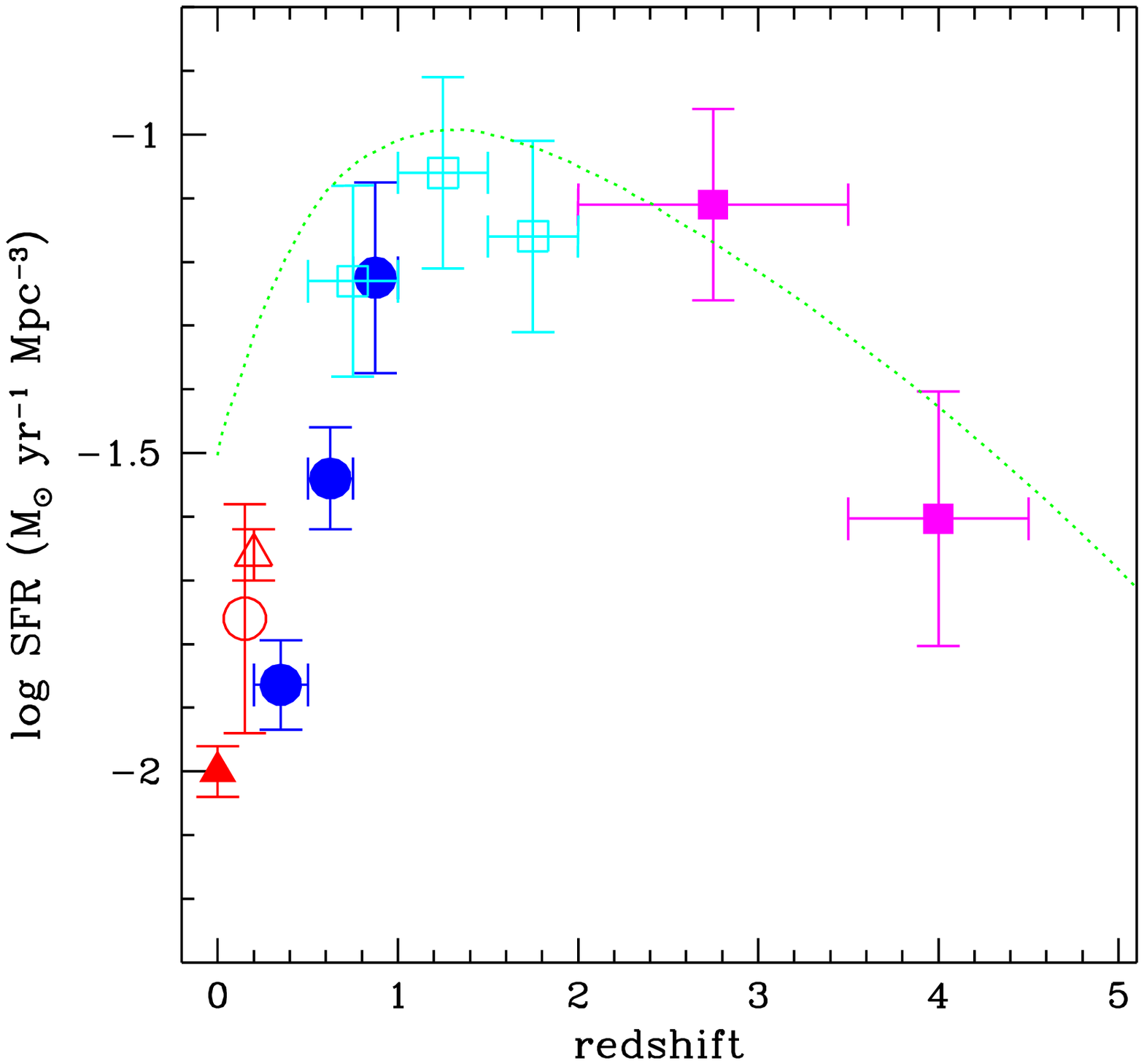,height=4.3in,width=4.5in}}
\caption{Cosmic star formation history. 
The data points with error bars are taken from the UV 
broadband measurements of Lilly \etal (1996) ({\it filled dots}), 
Connolly \etal (1997) ({\it empty squares}), Madau \etal (1996) and Madau 
(1997) ({\it filled squares}), Treyer \etal (1997) ({\it empty dot}).
Also plotted are the determinations from  the H$\alpha$ luminosity
functions of Gallego \etal (1995) ({\it filled triangle}) and 
Tresse \& Maddox (1997) ({\it empty triangle}). The data points have been 
converted to total SFR assuming a Salpeter IMF and a universal
$E(B-V)=0.1$. Because of dust extinction, the observed
rest-frame UV fluxes have been corrected upwards by a factor of 1.4 at 
2800 \AA\ and 2.1 at 1500 \AA. The {\it dotted line} depicts
the SFR per comoving volume predicted by a fiducial CDM model in which 
structure forms hierarchically (from Cole \etal 1994).  
\label{fig7}}
\end{figure}

\subsection{Star Formation at High Redshifts: Monolithic Collapse Versus 
Hierarchical Clustering Models}

The biggest uncertainty present in our estimates of the star formation 
density at $z>2$ is probably associated with dust reddening, but, as the 
color-selected HDF sample
includes only the most actively star-forming young objects, one could also
imagine the existence of a large population of relatively old or faint galaxies
still undetected at high-$z$. The issue of the amount of star formation at
early epochs is a non trivial one, as the two competing models, monolithic
collapse versus hierarchical clustering, make very different predictions in
this regard. From stellar population studies we know in fact that about half of
the present-day stars are contained in spheroidal systems, i.e., elliptical
galaxies and spiral galaxy bulges (Schechter \& Dressler 1987). In the
monolithic scenario these formed early and rapidly, experiencing a bright
starburst phase at high-$z$ (Eggen, Lynden-Bell, \& Sandage 1962; Tinsley \&
Gunn 1976). In hierarchical clustering theory instead ellipticals form 
continuosly by the merger of disk/bulge systems (Kauffman \etal 1993), and most
galaxies never experience star formation rates in excess of a few solar masses 
per year (Baugh \etal 1997). Figure 7 shows the star formation history 
predicted by a fiducial CDM model (normalized to reproduce the abundance of 
rich clusters, Cole \etal 1994) and compared to observational estimates.
Overall, the agreement between 
theoretical predictions and data is quite good. Both appear to 
produce only a small fraction, about $15-20\%$, of the current stellar 
content of galaxies at $z\gta 2-2.5$. In fact, the tendency to form 
the bulk of the stars at relatively low redshifts is a generic feature not 
only of the $\Omega_0=1$ CDM cosmology, but also of successful low-density 
CDM models (Cole \etal 1994; Baugh \etal 1997). While uncertainties still
remain in this comparison -- e.g., because of the poorly known effects of
dust obscuration and possible incompleteness in the high-$z$ sample -- one 
should note the tendency of the theoretical curve to sit above the data points 
at all epochs, thereby predicting luminosity-weighted mass-to-light ratios 
at the present time which are higher than observed. 

\begin{figure}
\centerline{\epsfig{file=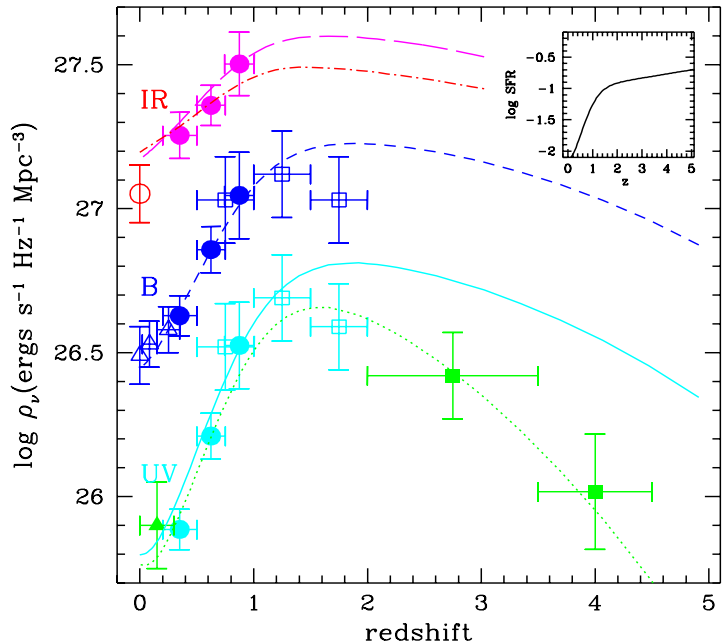,height=4.3in,width=4.5in}}
\caption{Test case with a much larger star formation density at high redshift
than indicated by the HDF dropout analysis. The model -- designed to mimick a
``monolithic collapse'' scenario -- assumes a Salpeter IMF and a dust opacity
which increases rapidly with redshift, $E(B-V)=0.011(1+z)^{2.2}$. Notation is
the same as in Figure 3. 
\label{fig8}}
\end{figure}

It is of interest then to ask how much larger could the volume-averaged SFR at
high-$z$ be before its fossil records -- in the form of long-lived, near
solar-mass stars -- became easily detectable as an excess of $K$-band light at
late epochs. In particular, is it possible to envisage a model where 50\%
of the present-day stars formed at $z>2.5$ and were shrouded by dust? The
predicted emission history from such a model is depicted in Figure 8. To
minimize the long-wavelength emissivity associated with the radiated
ultraviolet light, a Salpeter IMF has been adopted. Consistency with the HDF
data has been obtained assuming a dust extinction which increases rapidly with
redshift, $E(B-V)=0.011(1+z)^{2.2}$. This results in a correction to the rate
of star formation of a factor $\sim 5$ at $z=3$ and $\sim 15$ at $z=4$. The
total stellar mass density today is $\rho_s(0)=5.0\times 10^8\mdden$
($\Omega_sh_{50}^2=$0.007). 

Overall, the fit to the data is still acceptable, showing how the blue and
near-IR light at $z<1$ are {\it  relatively poor indicators of the star
formation history at early epochs}. The reason for this is the short timescale
available at $z\gta 2$, which makes the present-day stellar mass density rather
insensitive to a significant boost of the stellar birthrate at high redshifts.
By contrast, variations in the global SFR around $z\sim1.5$, where the
bulk of the stellar population was assembled, have a much larger impact.
The adopted extinction-redshift relation, in fact, implies negligible 
reddening at $z\lta 1$. Relaxing this -- likely unphysical -- assumption 
would cause the model to significantly overproduce the $K$-band local 
luminosity density. I have also checked that an even larger amount of hidden 
star formation at early epochs, as recently advocated by Rowan-Robinson \etal 
(1997) and Meurer \etal (1997), would generate too much blue, 1 \micron\, and 
2.2 \micron\ light to be still consistent with the observations. An IMF which 
is less rich in massive stars would only exacerbate the discrepancy. 

\subsection{The Ultraviolet Colors of Lyman-Break Galaxies}

\begin{figure}
\centerline{\epsfig{file=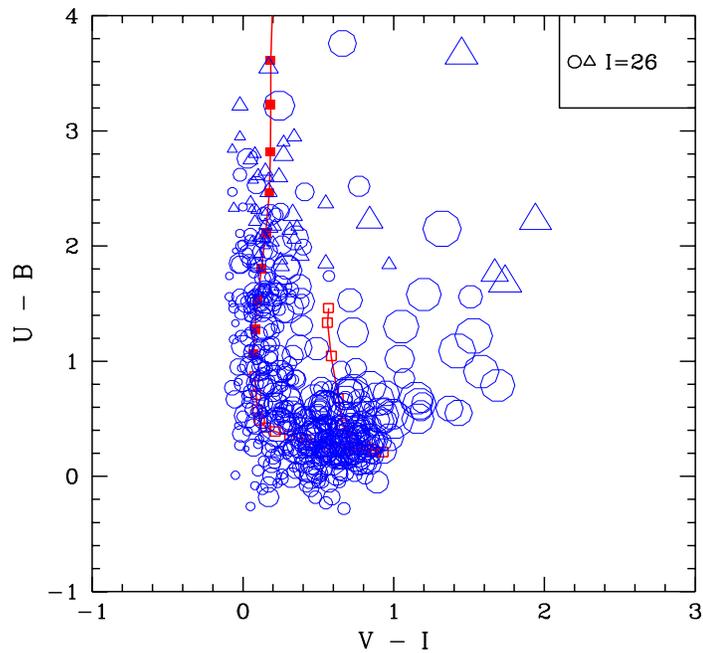,height=4.3in,width=4.5in}}
\caption{{\it Solid line:} model predictions for the color evolution of 
galaxies according to the star formation history of Figure 3. The points ({\it 
filled squares} for $z>2$ and {\it empty squares} for $z<2$) are plotted at 
redshift interval $\Delta z=0.1$. {\it Empty circles} and {\it triangles:} 
colors of galaxies in the HDF
with $22<B<27$. Objects undetected in $U$ (with $S/N<1$) are plotted as
triangles at the 1$\sigma$ lower limits to their $U-B$ colors. Symbols size
scales with the $I$ mag of the object, and all magnitudes are given in the 
AB system. The ``plume'' of Lyman-break galaxies is clearly seen in the
data. 
}
\label{fig9} 
\end{figure}

Figure 9 shows a comparison between the HDF data and the model predictions for
the evolution of galaxies in the $\ub$ vs. $\vi$ color-color plane according to
the star formation history of Figure 3. The HDF ultraviolet passband
-- which is bluer than the standard ground-based $U$ filter -- permits the 
identification of star-forming galaxies in the interval $2\lta z\lta 3.5$. 
Galaxies in this redshift range predominantly occupy the top left portion of
the $\ub$ vs. $\vi$ color-color diagram because of the attenuation by the 
intergalactic medium and intrinsic absorption (M96). Galaxies at lower redshift
can have similar $\ub$ colors, but are typically either old or dusty, and are
therefore red in $\vi$ as well. 

It is clear that the Salpeter IMF, $E(B-V)=0.1$ model reproduces quite well 
the rest-frame UV colors of high-$z$ objects in HDF. This may suggest that
interstellar dust is already present in these young galaxies and that it 
attenuates their 1500 \AA\ luminosities by a factor of $\sim 2$. A UV 
extinction of about 1 mag is also indicated by a comparison between
UV and H$\beta$ luminosities in three bright UV dropouts (Pettini \etal 
1997). One should note that the prescription for a ``correct'' de-reddening 
of the Lyman-break galaxies at $z\sim 3$ is the subject of an ongoing 
debate (e.g., Meurer \etal 1997; Dickinson \etal 1997; Pettini \etal 1997). 
Adopting the greyer extinction law deduced 
by Calzetti \etal (1994) from the integrated spectra of nearby starbursts 
would require larger corrections to the SFR at high-$z$ in order to match 
the observed colors. The consequence of this, however, would be the 
overproduction of red light at low redshifts, as noted in \S~5.4. Redder 
spectra can also result from an aging population or an IMF which is,
at early epochs, less rich in massive stars than the adopted ones.

\subsection{Constraints from the Mid- and Far-Infrared Background}

Ultimately, it should be possible to set some constraints on the total amount
of star formation that is hidden by dust over the entire history of the
universe by looking at the cosmic infrared background (CIB) (Fall \etal 1996;
Burigana \etal 1997; Guiderdoni \etal 1997).  Studies of the CIB provide
information which is complementary to that given by optical observations. If
most of the star formation activity takes place within dusty gas clouds, the
starlight which is absorbed by various dust components will be reradiated 
thermally at longer wavelengths according to characteristic IR spectra. The
energy in the CIB would then exceed by far the entire background optical light
which is recorded in the galaxy counts. 

From an analysis of the smoothness
of the {\it COBE} DIRBE maps, Kashlinsky \etal (1996) have recently set an
upper limit to the CIB of 10--15 nW m$^{-2}$ sr$^{-1}$ at $\lambda=$10--100
\micron\ assuming clustered sources which evolve according to typical
scenarios. An analysis using data from {\it COBE} FIRAS by Puget \etal (1996)
(see also Fixsen \etal 1996) has revealed an isotropic residual at a level of 3.4 ($\lambda/400
\micron)^{-3}$ nW m$^{-2}$ sr$^{-1}$ in the 400--1000 \micron\ range, which
could be the long-searched CIB. The detection, recently revisited by Guiderdoni
\etal (1997), should be regarded as uncertain since it depends critically on
the subtraction of foreground emission by interstellar dust. 

By comparison, the total amount of starlight that is absorbed by dust and
reprocessed in the infrared is 7.5 nW m$^{-2}$ sr$^{-1}$ in the model depicted
in Figure 3, about 30\% of the total radiated flux. The monolithic collapse
scenario of Figure 8 generates 6.5 nW m$^{-2}$ sr$^{-1}$ instead. The resulting
CIB spectrum is expected to be rather flat because of the spread in the dust
temperatures -- cool dust will likely dominate the long wavelength emission,
warm small grains will radiate mostly at shorter wavelengths --  and the 
distribution in redshift. While both models appear then to be
consistent with the data (given the large uncertainties associated with the 
removal of foreground emission and with the observed and predicted spectral
shape of the CIB), it is clear that too much infrared light would be generated
by scenarios that have significantly larger amount of hidden star formation at 
early and late epochs.

\subsection{Chemical Enrichment}

We may at this stage use our set of models to establish a cosmic timetable for 
the production of heavy elements (with atomic number $Z\ge 6$) in relatively 
bright
field galaxies (see M96). What we are interested in here is the universal rate
of ejection of newly synthesized material. In the approximation of
instantaneous recycling, the metal ejection rate per unit comoving volume can
be written as 
\begin{equation}
{\dot \rho_Z}=y(1-R)\times {\rm SFR}, \label{eq:rhoz}
\end{equation}
where the {\it net}, IMF-averaged yield of returned metals is 
\begin{equation}
y={\int mp_{\rm zm}\phi(m)dm\over (1-R)\int m\phi(m)dm},
\end{equation}
$p_{\rm zm}$ is the stellar yield, i.e., the mass fraction of a star of
mass $m$ that is converted to metals and ejected, and the dot denotes
differentiation with respect to cosmic time. 

The predicted end-products of stellar evolution, particularly from massive
stars, are subject to significant uncertainties. These are mainly due to the
effects of initial chemical composition, mass-loss history, the mechanisms of
supernova explosions, and the critical mass, $M_{\rm BH}$, above which stars
collapse to black holes without ejecting heavy elements into space (Maeder 1992;
Woosley and Weaver 1995). The IMF-averaged yield is also very sensitive to the 
choice of the IMF slope and lower-mass cutoff.
For a Scalo IMF in the assumed mass range ($0.1<M<125\msun$), for example, 
the net yield is typically a factor of 3.3 lower than Salpeter. At the same 
time, a lower cutoff of
$0.5\msun$ would boost the net yield by a factor of 1.9 for Salpeter and 1.7 
for Scalo. Note that some of these ambiguities partially 
cancel out when computing the total metal ejection rate, as the product
$y\times {\rm SFR}$ is less sensitive to the slope of the IMF than the yield or
the rate of star formation, and is insensitive to the lower mass cutoff.
Observationally, the best-fit ``effective yield'' (derived assuming a closed
box model) is 0.025$Z_\odot$ for Galactic halo clusters, 0.3$Z_\odot$ for disk
clusters, 0.4$Z_\odot$ for the solar neighborhood, and 1.8$Z_\odot$ for the
Galactic bulge (Pagel 1987). The last value may represent the universal true
yield, while the lower effective yields found in the other cases may be due,
e.g., to the loss of enriched material in galactic winds. 

\begin{figure}
\centerline{\epsfig{file=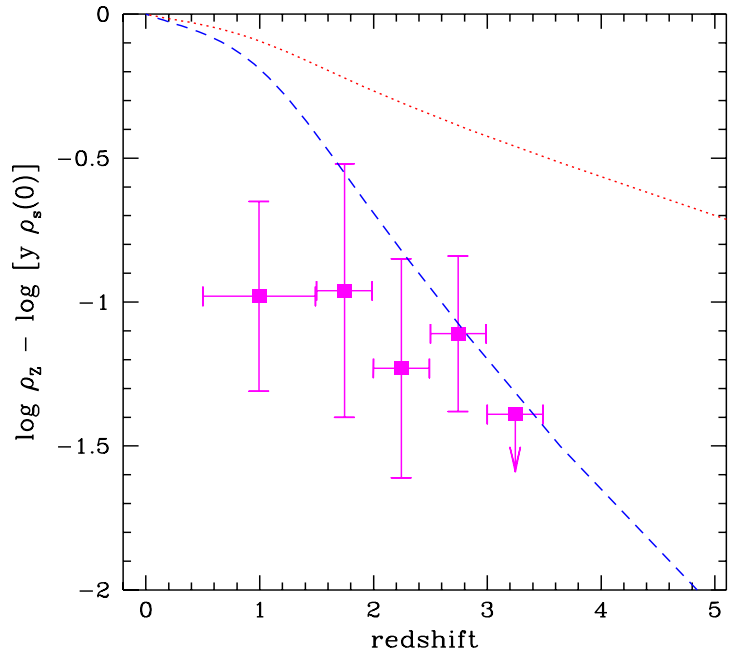,height=4.3in,width=4.5in}}
\caption{Total mass of heavy elements ever ejected versus redshift for the
Salpeter IMF model of Figure 3 ({\it dashed line}) and the monolithic
collapse model of Figure 8 ({\it dotted line}), normalized to $y\rho_s(0)$,
the total  mass density of metals at the present epoch. {\it Filled squares:}
column density-weighted metallicities (in units of solar) as derived from
observations of the damped Lyman-$\alpha$ systems (Pettini \etal 1997). 
\label{fig10}}
\end{figure}

Figure 10 shows the total mass of metals ever ejected, $\rho_Z$, versus 
redshift, i.e., the sum of the heavy elements stored in stars and in the gas
phase as given by the integral of equation (\ref{eq:rhoz}) over cosmic time.
The values plotted have been computed from the star formation histories
depicted in Figures 3 and 8, and have been normalized to $y\rho_s(0)$,
the mass density of metals at the present epoch according to
each model. A characteristic feature of the two competing scenarios is the
rather different average metallicity expected at high redshift. For comparison,
I have also plotted the {\it gas metallicity}, $Z_{\rm DLA}/Z_\odot$, as
deduced from observations by Pettini \etal (1997) of the damped Lyman-$\alpha$
systems (DLAs).  At early epochs,  when the gas consumption into stars is still
low, the metal mass density predicted from these models gives, in a closed box
model, a measurement of the metallicity of the gas phase. If DLAs and
star-forming field galaxies have the same level of heavy element enrichment,
then one would expect a rough agreement between $Z_{\rm DLA}$ and the model
predictions at $z\gta 3$. This is not true at $z\lta 2$, when a significant
fraction of heavy elements is locked into stars.\footnote{More complex 
chemical evolution models which reproduce the evolving gas content and the 
metal enrichment history of the DLAs have been developed by Lanzetta, Wolfe,
\& Turnshek (1995) and Pei \& Fall (1995).}~ From Figure 10, it does 
appear that the monolithic collapse model overpredicts the cosmic metallicity
at high redshifts as sampled by the DLAs. In order for such a model to be 
acceptable, the gas traced by the DLAs would have to be physically distinct
from the luminous star formation regions observed in the Lyman-break galaxies, 
and to be substantially under-enriched in metals compared to the cosmic mean.

\subsection{The Cluster-Field Analogy}

It has been recently pointed out by Renzini (1997) and Mushotzky \& Loewenstein
(1997) that, in the absence of any systematic cluster/field differences,
clusters of galaxies may also provide an indication of the metal formation
history of the universe. In the fiducial model of Figure 3, the global mean
metallicity of the local universe is $y\Omega_s/\Omega_b\sim 0.1
y/Z_\odot$ solar, to be compared with the overall cluster metal abundance,
$\sim 1/3$ solar. If $y\sim Z_\odot$, the efficiency of metal production must
have been larger in clusters than in the field, in spite of both having a
similar baryon-to-star conversion efficiency, $\Omega_s/\Omega_b\sim 10\%$. 
Cluster-related processes, like ram-pressure stripping, would
then be responsible for enriching the intracluster medium. Alternatively, a 
larger IMF-averaged metal yield, $y\sim 3 Z_\odot$, may solve the apparent 
discrepancy. In this case, field galaxies would have to have ejected most of 
the heavy elements they produce, and there should be a comparable share of 
metals in the intergalactic medium (IGM) as there is in the intracluster 
gas. The characteristic metal ejection rate
per unit comoving volume associated with such a pollution level would be large,
\begin{equation} 
{\dot \rho}_{Z,{\rm IGM}}\approx (0.013 \sfrd)
\left({\Omega_{\rm IGM}h_{50}^2\over 0.05}\right)\left({3 Z_{\rm IGM}\over
Z_\odot}\right) \left({f_{\rm inj}\over 0.7}\right)^{-1}
\left({\Delta t\over 2.5\, {\rm Gyr}}\right)^{-1}, 
\end{equation}
about ten times larger than derived at $z=2$ from our modeling of 
the galaxy emission history (with $y\sim Z_\odot$). Here, $\Omega_{\rm 
IGM}$ is the 
baryonic density parameter of the IGM phase and $f_{\rm inj}$ is the 
fraction of heavy elements injected into the IGM during a timescale $\Delta t$. 
It is hard to see, however, how massive galaxies with deep potential wells
could be responsible for large outflows of metal-enriched gas. Recent
observations of metal lines in the Ly$\alpha$ forest clouds at $z\sim 3$,
while pointing towards some widespread chemical enrichment at early epochs,  
suggest typical metallicities of only 0.003 to 0.01 solar (Cowie \etal 1995).
Similar low values have been inferred in local Ly$\alpha$ absorbers (Shull
\etal 1997). 

\acknowledgments

It is a pleasure to thank the hospitality of the European Southern 
Observatory, Garching, where this review was largely written.
I have benefited from many useful discussions on various topics related to this
talk with C. Baugh, G. Bruzual, S. Charlot, A. Connolly, M. Della Valle,  
M. Pettini, A. Renzini, M. Treyer, and my collaborators, L. Pozzetti and 
M. Dickinson. 
Support for this work was provided by NASA through grant AR-06337.10-94A from 
the Space Telescope Science Institute, which is operated by the Association 
of Universities for Research in Astronomy, Inc., under NASA contract NAS5-26555. 

\end{document}